# Multicomponent heat-bath configuration interaction with the perturbative correction for the calculation of protonic excited states

Naresh Alaal, Kurt R. Brorsen*

Department of Chemistry, University of Missouri, Columbia, MO, 65203, USA

*Email: brorsenk@missouri.edu

**Abstract**

In this study, we extend the multicomponent heat-bath configuration interaction (HCI) method to excited states. Previous multicomponent HCI studies have been performed using only the variational stage of the HCI algorithm as they have largely focused on the calculation of protonic densities. Because this study focuses on energetic quantities, a second-order perturbative correction after the variational stage is essential. Therefore, this study implements the second-order Epstein-Nesbet correction to the variational stage of multicomponent HCI for the first time. Additionally, this study introduces a new procedure for calculating reference excitation energies for multicomponent methods using the Fourier-grid Hamiltonian (FGH) method, which should allow the one-particle electronic basis set errors to be better isolated from errors arising from an incomplete description of electron-proton correlation. The excited-state multicomponent HCI method is benchmarked by computing protonic excitations of the HCN and FHF$^-$ molecules and is shown to be of similar accuracy to previous excited-state multicomponent methods such as multicomponent time-dependent density functional theory and equation-of-motion coupled-cluster theory relative to the new FGH reference values.

## I. Introduction

Owing to its huge practical relevance in the interpretation of experimental data, the more accurate and more efficient calculation of electronic[1-3] and vibrational excited states[4-7] is a long-standing goal in quantum chemistry. The most common approach for the calculation of electronic excited states of single-reference systems is probably based on a response framework either using time-dependent density functional theory (TDDFT)[8-10] or, for smaller chemical systems, coupled-cluster theory,[11-12] while for vibrational excited states many-body,[13-18] configuration interaction (CI)[19-21] or related approaches[22-26] are common.

In the last few years,[27-35] selected CI approaches[36-38] that were originally developed for the calculation of ground-state electronic energies of strongly correlated systems have been extended to the calculation of excited states.[39-41] These selected CI approaches approximate a full CI or complete active space CI (CASCI) calculation by only selecting a subset of the Slater determinants from the full CI wave function to include in the wave function expansion using a rule or selection criterion, which keeps the calculation tractable. After this so-called variational stage of the calculation has converged, the contribution of the remaining Slater determinants are included perturbatively, which is typically done using second-order Epstein-Nesbet perturbation theory (PT2)[42] as this facilitates an efficient implementation. As has been consistently shown, the PT2 correction is essential to include for accurate energetic calculations and the PT2 correction also allows for various extrapolation approaches to estimate the full CI or CASCI limit.[27, 29] Selected CI + PT2 approaches have proven to be highly accurate for the calculation of excited states relative to other state-of-the-art *ab initio* methods.[39, 41]

Selected CI approaches have also been widely used for the calculation of vibrational excited states.[43-49] More recently, the progress in selected CI for the calculation of electronic excited states has been adapted to the calculation of vibrational excited states[50-52] and the resulting methods have shown state-of-the-art performance for medium- and large-sized systems. For the vibrational methods based on the recent electronic selected CI methods, the PT2 correction has proven to be essential. This is especially true for larger systems such as naphthalene where the inclusion of the PT2 correction can reduce the root-mean-squared error of the calculated vibrational excitation energies by over a factor of three.[52]

Excited states can also be computed[53-56] using modern multicomponent methods,[57-63] which treat select nuclei, typically protons, quantum mechanically and identically to electrons. Multicomponent methods differ from other quantum mechanical methods for computing excited states in that there are two fundamentally different types of excited states: (1) electronic excited states, which are similar in nature to the excited states in standard or single-component quantum chemistry and (2) nuclear or protonic excited states, which corresponds to a vibrational-like excitation of the quantum nuclei. As the electronic and protonic systems are coupled, an excitation of one subsystem causes the other system to change during these excitations, but, in practice, most low-lying excitations are dominated by a single type of excitation. We do note that there also exist mixed electron-protonic excited states,[55] where both an electron and proton have been excited, though we do not focus on them in this study.

Compared to the computation of electronic or vibrational excited states, computing multicomponent excited states for small molecules with even qualitatively accuracy is a relatively recent achievement with a multicomponent TDDFT study being the first in 2018.[53-54] More recently, the multicomponent EOM-CCSD method has been introduced[55-56] and has been shown to be of comparable accuracy to multicomponent TDDFT. While both methods represent huge steps forward for the application of multicomponent methods, they can still differ from benchmark Fourier-grid Hamiltonian (FGH) calculations[64-65] by hundreds of wavenumbers, which potentially reduces their ability to be used in applications.

An additional difficulty in the development of excited-state multicomponent methods is disentangling the one-electron basis set error from the correlation energy error. For calculations that treat only a single proton quantum mechanically, FGH calculations are relatively insensitive to the electronic basis set. As an example, an FGH calculation on the HCN molecule with the hydrogen atom treated quantum mechanically gives a first protonic excitation energy of 683 $cm^{-1}$ when the cc-pVDZ basis set is used for all atoms and a first protonic excitation energy of 710 $cm^{-1}$ when the cc-pV5Z basis set is used for the H atom and the cc-pVDZ basis set is used for the C and N atoms. This differs materially from multicomponent excited-state calculations where the analogous multicomponent TDDFT calculation[53] with the epc17-2 electron-proton correlation functional[66-67] gives a first protonic excitation energy of 2818 $cm^{-1}$ or 820 $cm^{-1}$ with either the cc-pVDZ or cc-pV5Z electronic basis set on the hydrogen atom, respectively. While we focus on protonic excitation energies in this study, we note that the dependence on the one-particle basis centered on the quantum nuclei also occurs for other protonic properties such as the protonic density as was discussed in a recent multicomponent density-matrix renormalization group study.[68-69]

The need for large electronic basis sets centered on the hydrogen atom to reproduce reference FGH calculations makes the introduction of new multicomponent methods difficult as the implementation must typically be efficient enough to handle more than 100 electronic basis functions, which means considerable time and effort must be devoted to the computational implementation even before the performance of the method is well understood. This can significantly hinder the rapid development of new multicomponent methods. As an example, the parallel implementation of the multicomponent reduced explicitly-correlated Hartree-Fock method[70-71] has been run with over 8,000 cores, but it was only after the parallel implementation was finished was it possible to determine that the RXCHF method could not qualitatively describe the protonic density of three atom systems such as FHF⁻ or HCN. A part of the goal of this study is to introduce a new procedure for generating reference FGH calculations. This should reduce the programmer effort required for new multicomponent methods such that only after it is understood whether a new multicomponent is likely to perform well will it be necessary to optimize the computational implementation.

This difference in behavior between the FGH and multicomponent methods has been attributed[53-54] to the fact that when the grid is constructed in the FGH calculations, the electronic basis set on the quantized particle follows the position of the point charge nucleus, which significantly increases the flexibility of the one-particle electronic basis set relative to the multicomponent calculations where the electronic basis set is fixed at a single point in space. To allow a better comparison between FGH and multicomponent calculations, we introduce a new procedure for the FGH calculations that has the same basis set limitations as their multicomponent counterparts and thereby allow us to better benchmark multicomponent methods without the need for computationally expensive calculations with large electronic basis sets centered on the quantum nuclei. This is similar in spirit to how single-component methods are benchmarked. A recent example is a paper[72] that calculated the ground-state energy of the benzene molecule with a variety of state-of-the-art single-component *ab initio* methods with the cc-pVDZ electronic basis set. Such calculations are not close to the complete basis set limit, but the study is still useful for providing detailed information about how well the various methods describe the electron correlation energy for a given one-particle basis set.

The second goal of this study is to extend the multicomponent selected CI method, multicomponent heat-bath configuration interaction (HCI),[73-74] to the calculation of excited states by modifying the selection criterion to choose important multicomponent configurations for both the ground and select excited states. As accurate excited states in modern selected-CI methods require the PT2 correction, we also introduce a multicomponent PT2 correction to multicomponent HCI for the first time. We show that the multicomponent HCI with the PT2 correction reproduces the new benchmarking FGH calculations well for the HCN and FHF⁻ molecules and is of similar accuracy compared to other state-of-the-art multicomponent methods for the calculations of protonic excitation energies.

The paper is organized as follows. In Section II, we introduce an FGH procedure appropriate for benchmarking multicomponent protonic excitation energies and validate the approach by comparing the results to existing excited-state multicomponent methods. In Section III, we introduce the multicomponent HCI method, extend the method to the computation of excited state, and derive and implement the multicomponent PT2 correction for multicomponent HCI. In Section IV, we benchmark the multicomponent HCI method for the calculation of protonic excited states. In Section V, we conclude and discuss the prospects of the multicomponent HCI method.

## II. Reference Fourier-grid Hamiltonian Excited-State Energies

Single-component quantum chemistry calculations using Gaussian-type orbitals basis functions are almost always performed with the electronic basis functions centered on the atomic nuclei. This has been true for all previous FGH benchmark calculations of multicomponent methods. In these FGH calculations, a grid potential is constructed from a series of single-component quantum chemistry calculations where the particle that is being quantized (typically a proton nucleus as in this study) is moved with the positions of all other nuclei held fixed. In multicomponent methods, the quantized nuclei or particle is no longer a point particle fixed at a single point in space as the Born-Oppenheimer approximation is no longer assumed. However, the electronic basis set is fixed at a preselected point in space so that it cannot follow the quantum nuclei as in the FGH calculations. As has been previously discussed,[53-54] this functionally causes the electronic basis sets in FGH and multicomponent calculations to not be of the same quality.

To better align the quality of the electronic basis sets in FGH and multicomponent calculations, a simple change is to fix the position of the electronic basis set during the FGH calculations and allow the nucleus to move as a bare point charge as this restriction is essentially what occurs within the multicomponent framework. Such an FGH calculation is simple to implement in most electronic structure theory codes using ghost atoms to fix the position of the electronic basis functions and a point charge with no electronic basis functions centered on it for the nucleus in the individual *ab initio* calculations during the FGH process. This procedure will invariably reduce the accuracy of the FGH calculation for a given electronic basis set relative to the complete basis set limit, but it will also facilitate the benchmarking of multicomponent methods for the inclusion of electron-proton correlation energy and the calculation of protonic properties for a chosen electronic basis set, which should speed up the development of new multicomponent methods and make the comparison of the accuracy of multicomponent methods easier.

We have performed a series of FGH calculations on the $FHF^-$ and HCN molecules with the hydrogen proton treated quantum mechanically with a variety of different electronic basis set combinations for the hydrogen and all other heavy atoms. Two sets of FGH calculations with either single-component CCSD or single-component density-functional theory with the B3LYP exchange-correlation functional[75] were performed. All FGH calculations used a grid of 32 points in each dimension with an initial value of -0.5806 Å and a final value of 0.6194 Å. FGH calculations were performed in two ways. First, the hydrogen electronic basis functions were allowed to follow the proton as has been standard in FGH calculations. These calculations are labeled "Move." Secondly, calculations were performed with the hydrogen atom basis functions fixed at the origin as the protonic point charge was allowed to move on the grid. These later calculations are labeled "Fixed." For all FGH calculations, the positions of the non-hydrogen atoms were determined from a geometry optimization at the CCSD(T)/cc-pV5Z level of theory with the hydrogen atom fixed at the origin. These optimizations were performed with CFOUR.[76]

Results from the FGH calculations as well as previous multicomponent TDDFT and EOM-CCSD protonic excitation energies can be found in Table 1. The multicomponent TDDFT calculations were performed with an even-tempered[77] $8s8p8d$ protonic basis set with the B3LYP exchange-correlation functional[75] and epc17-2 electron-proton correlation functional[66-67] while the multicomponent EOM-CCSD calculations were performed with the PB4F-2 protonic basis set.[78] For both the HCN and FHF[-] molecules, the first and second protonic excited states are doubly degenerate and are normally referred to as "bending" modes as the excited-state protonic wave functions are mostly localized perpendicular to a line drawn through the position of the classical nuclei. The third excited state is singly degenerate and called a "stretching" mode as it is mostly localized parallel to a line drawn through the position of the classical nuclei.

**HCN**

| | CCSD | | | | | | DFT | | | | | |
| | Fixed | | Move | | EOM-CCSD | | Fixed | | Move | | TDDFT[53] | |
| n_state | 4 | 2,3 | 4 | 2,3 | 4 | 2,3 | 4 | 2,3 | 4 | 2,3 | 4 | 2,3 |
| DZ/DZ | 4231 | 2935 | 3104 | 683 | 4375 | 3028 | 4194 | 2858 | 3087 | 699 | 4208 | 2818 |
| TZ/DZ | 3440 | 1928 | 3104 | 695 | 4038 | 2282 | 3402 | 1921 | 3087 | 704 | 3851 | 2229 |
| QZ/DZ | 3202 | 1076 | 3144 | 706 | 3738 | 1756 | 3210 | 1055 | 3100 | 706 | 3376 | 1362 |
| 5Z/DZ | 3148 | 899 | 3152 | 719 | | | 3131 | 882 | 3092 | 712 | 3217 | 820 |
| TZ/TZ | 3275 | 1845 | 3100 | 685 | | | 3275 | 1858 | 3074 | 690 | 3731 | 2164 |
| QZ/TZ | 3219 | 1045 | 3114 | 686 | | | 3207 | 1040 | 3079 | 692 | 3351 | 1350 |
| 5Z/TZ | 3154 | 876 | 3117 | 691 | | | 3134 | 868 | 3076 | 693 | 3216 | 924 |

**FHF⁻**

| | CCSD | | | | | | DFT | | | | | |
| | Fixed | | Move | | EOM-CCSD | | Fixed | | Move | | TDDFT | |
| n_state | 4 | 2,3 | 4 | 2,3 | 4 | 2,3 | 4 | 2,3 | 4 | 2,3 | 4 | 2,3 |
| DZ/DZ | 3817 | 2747 | 2193 | 1289 | 3775 | 2755 | 3862 | 2770 | 2313 | 1139 | 3331 | 2653 |
| TZ/DZ | 3030 | 2197 | 2071 | 1390 | 2980 | 2309 | 3111 | 2174 | 2243 | 1230 | 2498 | 2294 |
| QZ/DZ | 2458 | 1587 | 2072 | 1465 | 2135 | 1914 | 2547 | 1539 | 2289 | 1277 | 2001 | 1618 |
| 5Z/DZ | 2272 | 1426 | 1991 | 1492 | | | 2347 | 1364 | 2292 | 1270 | 1828 | 1401 |
| TZ/TZ | 2916 | 2082 | 2097 | 1290 | | | 3033 | 2063 | 2211 | 1172 | 2414 | 2128 |
| QZ/TZ | 2355 | 1495 | 2044 | 1307 | | | 2466 | 1432 | 2185 | 1185 | 1924 | 1529 |

**Table 1**: Protonic excitation energies calculated with the FGH, multicomponent TDDFT, and multicomponent EOM-CCSD methods for the HCN and FHF⁻ systems. All excitation energies in $cm^{-1}$ and are relative to the ground-state energy (n_state = 1). An n_state of 2 or 3 corresponds to a degenerate bending mode and an n_state of 4 corresponds to the stretching mode. A basis set combination is denoted as XZ/YZ where "X" can be D, T, Q… and indicates the zeta-level of the hydrogen atom electronic basis set. "Y" is defined similarly, but for all non-hydrogen atoms.

As the results in Table 1 demonstrate, the multicomponent TDDFT and EOM-CCSD calculations agree much better with FGH calculations where the hydrogen electronic basis functions are fixed at the optimized position of the hydrogen atom. Given the results here, we suggest that fixed FGH calculations should be used for initial benchmarking of multicomponent methods as they allow the one-particle electronic basis set error to be better isolated such that the relative performance of multicomponent methods for including electron-proton correlation can be better assessed. As predicted, the move FGH calculations are much better converged to the complete basis set limit with respect to the basis set on the hydrogen atom. Ultimately, for quantitatively accurate calculations, it is evident that when using current electronic basis sets, large electronic basis sets are needed for the hydrogen atom, but the results presented here should facilitate the development of new methods as prototype implementations no longer need to be able to computationally tractable with 5z electronic basis sets in order to compare to benchmarking data.

### III. Multicomponent Heat-Bath Configuration Interaction

In addition to the calculation of more applicable reference FGH protonic excitation energies, the second goal of this study is to extend the multicomponent HCI method[73-74] to the calculation of protonic excitation energies. We first briefly discuss the multicomponent HCI method before introducing the modification of the selection criterion appropriate for the calculation of protonic excited states and the PT2 correction for each of these states. We restrict the discussion to a single proton treated quantum mechanically, but generalization to multiple quantum nuclei is straightforward. We adopt the convention that the letters *i, j, k…* denote occupied electronic orbitals, *a, b, c…* denote unoccupied electronic orbitals, and *p, q, r…* denote any type of electronic orbitals. Protonic orbitals are defined analogously, but with capital letters. Greek letters are used to index multicomponent configurations.

In multicomponent HCI, the wave function, $|\Psi\rangle$, is written as a sum of multicomponent configurations, $|\Psi_{\mu\nu}\rangle$, as

$$|\Psi\rangle = \sum_{\mu\nu} C_{\mu\nu} |\Psi_{\mu\nu}\rangle. \tag{1}$$

A multicomponent configuration is written as a direct product of electronic and protonic Slater determinants as

$$|\Psi_{\mu\nu}\rangle = |\Psi_{\mu}^{\text{elec}}\rangle \otimes |\Psi_{\nu}^{\text{prot}}\rangle, \tag{2}$$

where "elec" and "prot" denote the electronic and protonic Slater determinants, respectively. The sum in Eq. 1 is over a selected subset of multicomponent configurations from the full CI space, which is called the variational space.

HCI calculations normally consist of two stages:[29-30, 33, 39] (1) a variational stage in which the variational space for the sum in Eq. 1 is found and the Hamiltonian matrix is constructed and then diagonalized to calculate a variational ground-state energy and (2) a perturbative stage where the impact of multicomponent configurations not in the variational space is included using second-order Epstein-Nesbet perturbation theory. Previous multicomponent HCI calculations have only implemented the variational stage.[73-74]

The variational stage begins with the selection of an initial variational space, which is typically a single multicomponent configuration, but, in principle, it could be a small

CASCI space. From this initial variational space, a multicomponent configuration, $|\Psi_{\lambda\sigma}\rangle$, not in the current variational space is added to it if

$$\left|H_{\mu\nu,\lambda\sigma}C_{\mu\nu}\right| > \varepsilon_{\text{VAR}}, \tag{3}$$

where $|\Psi_{\mu\nu}\rangle$ is a multicomponent configuration currently in the variational space and $\varepsilon_{\text{VAR}}$ is a user-chosen cutoff value. $H_{\mu\nu,\lambda\sigma}$ is an element of the multicomponent Hamiltonian matrix. The multicomponent Hamiltonian, *H*, excluding classical nuclear repulsion with a single-quantum proton can be written in second-quantized form as

$$H = \sum_{pq} h_{pq}^{\text{elec}}\hat{E}_{pq} + \sum_{PQ} h_{PQ}^{\text{prot}}\hat{E}_{PQ} + \frac{1}{2}\sum_{pqrs}(pq|rs)\hat{E}_{pq,rs} - \sum_{pqPQ}(pq|PQ)\hat{E}_{pq,PQ} \tag{4}$$

where $\hat{E}_{pq}$, $\hat{E}_{PQ}$, $\hat{E}_{pq,rs}$, and $\hat{E}_{pq,PQ}$ are the one-particle electronic, one-particle protonic, two-particle electronic, and two-particle electronic-protonic excitation operators, respectively.[74] Using the enlarged variational space, the Hamiltonian is constructed and diagonalized to find the ground-state energy and HCI expansion coefficients in Eq. 1. The variational space is then once again enlarged using the selection criterion in Eq. 3 and the iterations of the variational stage are repeated until either the change in energy between iterations or the number of multicomponent configurations added to the variational space falls below a user-defined value. The advantage of multicomponent HCI is that the selection criterion in Eq. 3 facilitates the use of a look-up dictionary for determining which multicomponent configurations should be added to the variational space, which greatly increases computational efficiency. More details can be found in the previous multicomponent HCI studies.[73-74]

It is possible to calculate the energy of HCI expansion coefficients of more than a single state when diagonalizing the Hamiltonian in the variational space basis. However, if the selection criterion in Eq. 3 is used, the variational space is chosen based on the structure of the ground-state wave function and the computed excited-state energies and coefficients are poor. Therefore, to extend the multicomponent HCI method to the calculation of excited states, we modify the selection criterion in a manner similar to previous HCI studies.[39, 41, 51-52] For a multicomponent HCI calculation of the lowest $N$ states, at the start of each iteration during the variational stage, we calculate, the vector $\boldsymbol{C}^{\text{MAX}}$ with its elements defined as

$$C_{\mu\nu}^{\text{MAX}} = \max\left(\left|C_{\mu\nu}^1\right|, \left|C_{\mu\nu}^2\right|, \dots, \left|C_{\mu\nu}^N\right|\right) \tag{5}$$

where $C_{\mu\nu}^i$ is the expansion coefficient of state $i$. The vector $\boldsymbol{C}^{\text{MAX}}$ is then used to enlarge the variational space using the selection criterion in Eq. 3 with the rest of the variational stage preceding identically to a ground-state calculation.

Once the variational stage has been completed, we correct each of the states using second-order Epstein-Nesbet perturbation theory. In this theory, the reference Hamiltonian, $H_0$ is written as

$$H_0 = \sum_{\mu\nu,\lambda\sigma}^{\text{variational}} H_{\mu\nu,\lambda\sigma} |\Psi_{\mu\nu}\rangle\langle\Psi_{\lambda\sigma}| + \sum_{\lambda\sigma}^{\text{external}} H_{\lambda\sigma,\lambda\sigma} |\Psi_{\lambda\sigma}\rangle\langle\Psi_{\lambda\sigma}| \tag{6}$$

which leads to a PT2 correction for the $i$th state of

$$E_2^i = \sum_{\lambda\sigma}^{\text{external}} \frac{\left(\sum_{\mu\nu}^{\text{variational}} H_{\mu\nu,\lambda\sigma} C_{\mu\nu}^i\right)^2}{E_0 - H_{\lambda\sigma,\lambda\sigma}} \tag{7}$$

where "variational" and "external" indicate that the sum is over all multicomponent configurations in the final and not in the final variational space, respectively. In practice, the sum over the external space is computationally intractable as it involves a huge number of multicomponent configurations and so the sum is restricted to only include contributions where

$$\left| H_{\mu\nu,\lambda\sigma} C_{\mu\nu}^{i} \right| > \varepsilon_{\mathrm{PT2}}, \tag{8}$$

where $\varepsilon_{\mathrm{PT2}}$ is a user-defined cutoff value. This allows the dictionary for enlarging the variational space to be used for screening, which greatly limits the number of terms that must be calculated in Eq. 7. Additional details about the structure of this dictionary can be found in the original multicomponent HCI study.[73] We note that that in single-component HCI, the original PT2 implementation has been modified[30, 33] to be computed stochastically and semi-stochastically as this increases computational efficiency and allows the PT2 correction to be applied with smaller values of $\varepsilon_{\mathrm{PT2}}$. In principle, this could be done for multicomponent HCI as well, but in this study, we focus on the deterministic PT2 correction. As will be discussed more in Section IV, the implementation of the PT2 correction also allows various extrapolation approaches to be used to estimate the full CI or, equivalently, the $\varepsilon_{\mathrm{VAR}} \to 0$ limit.

All the methods in this study were implemented in a locally modified predecessor of the Arrow code,[29-30, 33] which was originally designed for single-component HCI.

## IV. Computational Methods

We have benchmarked the multicomponent HCI method on the HCN and FHF⁻ molecules. While the single-component HCI method typically seeks to approximate a full CI or CASCI wave function, as discussed in the previous multicomponent HCI study,[73] such an approach can make it difficult to include electron-proton correlation rather than electron-electron correlation without using very small values of $\varepsilon_{VAR}$  Therefore, similar to the previous multicomponent HCI study, we use the HCI selection algorithm to approximate a truncated CI wave function rather than a full CI or CASCI wave function, which for this study is limited to a maximum of triple electronic excitation with single protonic excitations. We briefly mention that the computational scaling with respect to system size of multicomponent HCI is formally equal to that of the underlying multicomponent HCI method because the exact multicomponent CI answer is obtained in the limit that $\varepsilon_{VAR} \rightarrow 0$. As we seek to approximate a truncated CI method with up to quadruple excitations, the multicomponent HCI method in this study computationally scales $N^{10}$ with respect to system size. The multicomponent EOM-CCSD method mentioned previously scales $N^6$ with respect to system size.

For all multicomponent HCI calculations, the proton was treated quantum mechanically. The geometries of the systems were obtained from single-component CCSD geometry optimizations using the aug-cc-pVTZ electronic basis set[79-80] for all atoms. All multicomponent HCI calculations used the PB4F-2 protonic basis set.[78] Similar to the benchmarking FGH calculations, we use multiple combinations of the correlation-consistent electronic basis sets. A combination is denoted as XZ/YZ where "X" can be D, T, Q… and indicates the zeta-level of the hydrogen atom electronic basis set. "Y" is defined identically, but for all non-hydrogen atoms. As detailed in Section V, we tested a range of $\varepsilon_{VAR}$ and $\varepsilon_{PT2}$ values to investigate the convergence of the multicomponent HCI algorithm and to perform the extrapolation procedure.

For both test systems, the four states of lowest energy were found, for a total of three excited states, which is identical to previous multicomponent TDDFT studies. The initial guess wave function for each state consists of a multicomponent configuration with electronic multicomponent HF orbitals and ground- or excited-state multicomponent HF protonic orbitals. For all calculations, we number our states starting at 1, *i.e.*, the ground state is state 1. All excited state energies are expressed relative to the ground-state.

## V. Results and Discussion

## A. Energy Convergence

We first test the convergence of the PT2 correction energy with respect to $\varepsilon_{\mathrm{PT2}}$ for ground- and excited-states. For all calculations, we use an $\varepsilon_{\mathrm{VAR}}$ of $1\times10^{-4}$. Results from the ground-state calculation with the QZ/QZ electronic basis set are shown in Figure 1. We plot total energies, *i.e.*, combined variational and PT2 energies. As expected, for both the systems, total energies decrease as a function of $\varepsilon_{\mathrm{PT2}}$ with convergence achieved as $\varepsilon_{\mathrm{PT2}}$ is decreased below $1\times10^{-6}$ for both the HCN and FHF$^-$ molecules. Though it is not shown here, the convergence with other combinations of electronic basis sets is similar.

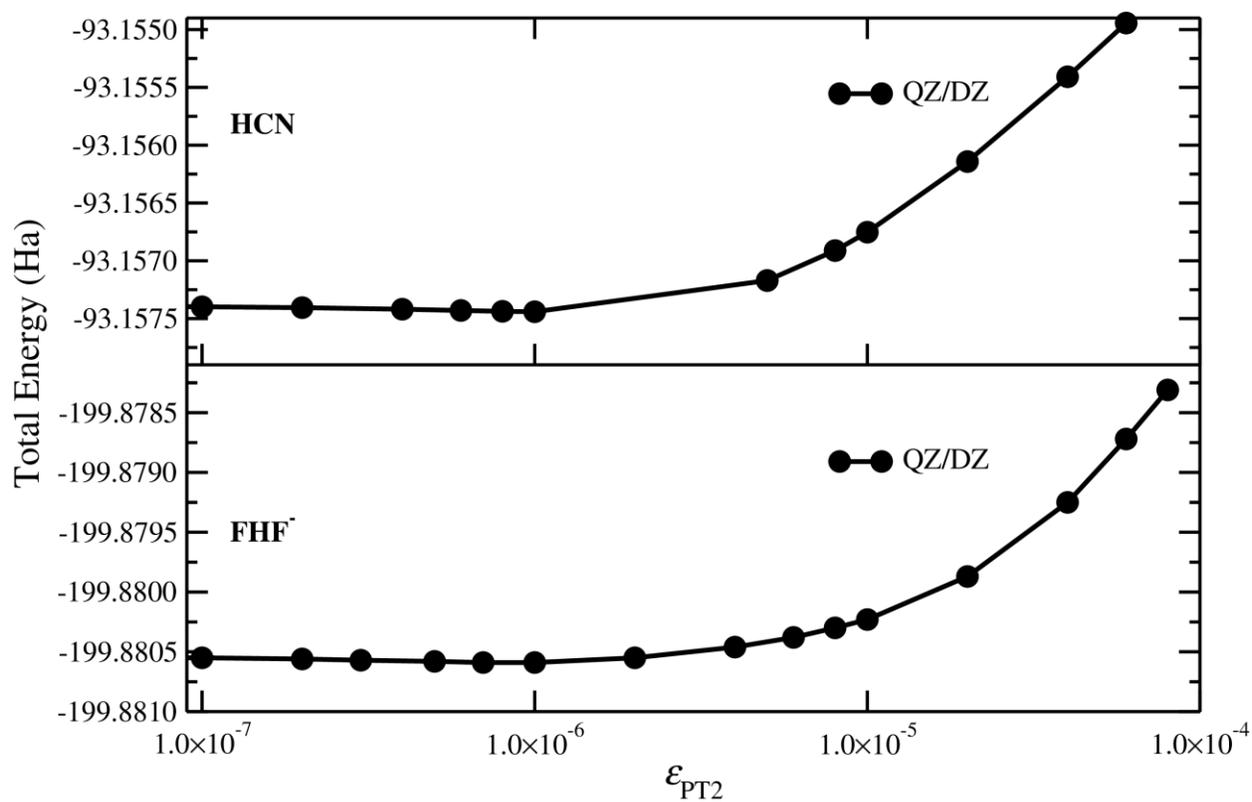

Figure 1: Plot of the total multicomponent HCI energy of the HCN (top) and FHF⁻ (bottom) molecules as a function of $\varepsilon_{PT2}$. The smallest $\varepsilon_{PT2}$ value is $1 \times 10^{-7}$. Results are shown for the QZ/DZ electronic basis set combination. $\varepsilon_{VAR}$ was set to $1 \times 10^{-4}$ for all calculations.

Next, we examine the convergence of the PT2 correction for the excited-state multicomponent calculations. Once again, for all calculations, we use an $\varepsilon_{\mathrm{VAR}}$ of 1x10$^{-4}$. Results from these calculations are shown in Figures 2 and 3. Both systems once again achieve convergence for values of $\varepsilon_{\mathrm{PT2}}$ below 1x10$^{-6}$. All states show similar trends in their total energy with respect to $\varepsilon_{\mathrm{PT2}}$. From the results in Figures 1-3, we are confident that we can converge the perturbative correction for the systems in this study.

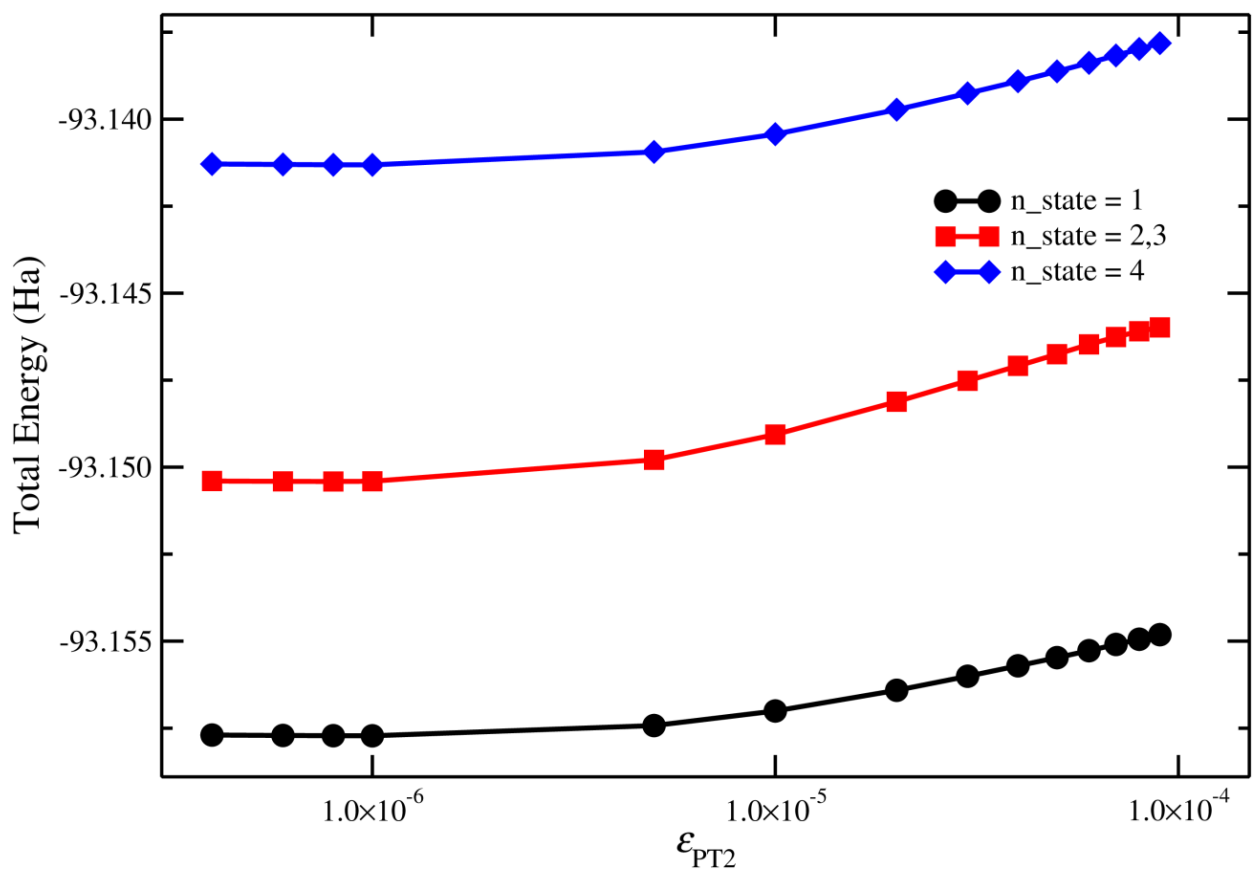

**Figure 2:** Plot of the total multicomponent HCI energy of HCN for the ground- and excited-states for different values of $\varepsilon_{PT2}$. The smallest $\varepsilon_{PT2}$ value considered here is $4.0 \times 10^{-7}$. All calculations in this plot used the QZ/DZ electronic basis set combination. $\varepsilon_{VAR}$ was equal to $1 \times 10^{-4}$ for all calculations. An n_state of 2 or 3 corresponds to a degenerate bending mode and an n_state of 4 corresponds to the stretching mode.

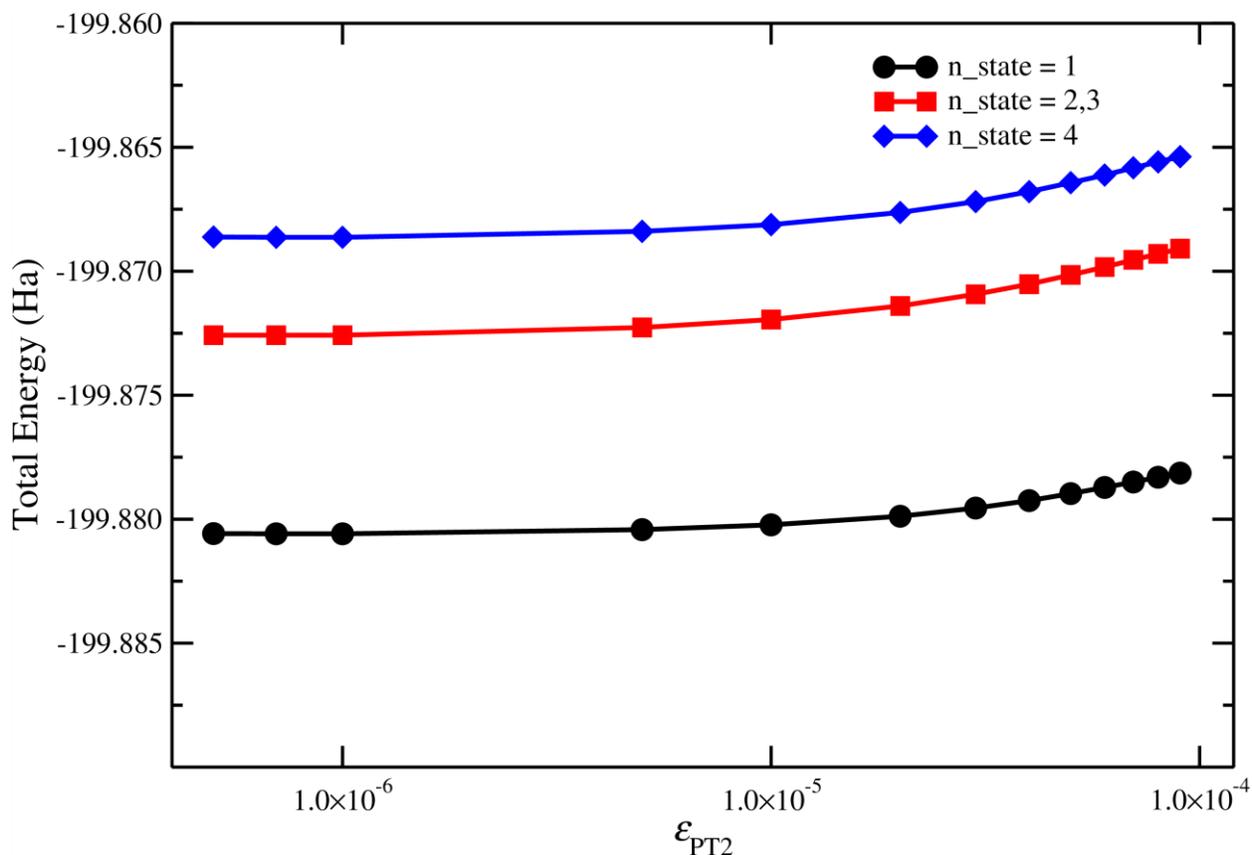

**Figure 3:** Plot of the total multicomponent HCI energy of FHF⁻ for the ground- and excited-states for different values of $\varepsilon_{PT2}$. The lowest $\varepsilon_{PT2}$ value considered here is $5.0 \times 10^{-7}$. All calculations in this plot used the QZ/DZ electronic basis set combination. $\varepsilon_{VAR}$ was equal to $1 \times 10^{-4}$ for all calculations. An n_state of 2 or 3 corresponds to a degenerate bending mode and an n_state of 4 corresponds to the stretching mode.

**B. Excited State Energies**

Next, we discuss the calculated multicomponent HCI excited-state energies of the HCN and FHF⁻ molecules. We have performed the calculations with the DZ/DZ, TZ/DZ and QZ/DZ electronic basis set combinations. We calculate the four lowest-energy states. We use the smallest possible values of $\varepsilon_{\mathrm{VAR}}$ and $\varepsilon_{\mathrm{PT2}}$ for each electronic basis set combination given our implementation and computer resources. These values and the results from the multicomponent HCI calculation are shown in Table 2.

| HCN | | | | | | | |
|---|---|---|---|---|---|---|---|
| | | HCI | | | | FGH | |
| | $\varepsilon_{VAR}$ | $\varepsilon_{PT2}$ | Var | | Var + PT2 | | | |
| State | | | 4 | 2,3 | 4 | 2,3 | 4 | 2,3 |
| DZ/DZ | $7.5 \times 10^{-6}$ | $5 \times 10^{-8}$ | 4277 | 2972 | 4274 | 2968 | 4231 | 2935 |
| TZ/DZ | $1.5 \times 10^{-5}$ | $2 \times 10^{-7}$ | 3867 | 2165 | 3847 | 2147 | 3440 | 1928 |
| QZ/DZ | $3 \times 10^{-5}$ | $4 \times 10^{-7}$ | 3615 | 1686 | 3575 | 1621 | 3202 | 1076 |

| FHF- | | | | | | | |
|---|---|---|---|---|---|---|---|
| | | HCI | | | | FGH | |
| | $\varepsilon_{VAR}$ | $\varepsilon_{PT2}$ | Var | | Var + PT2 | | | |
| State | | | 4 | 2,3 | 4 | 2,3 | 4 | 2,3 |
| DZ/DZ | $8 \times 10^{-6}$ | $3 \times 10^{-7}$ | 3843 | 2772 | 3842 | 2772 | 3817 | 2747 |
| TZ/DZ | $3 \times 10^{-5}$ | $4 \times 10^{-7}$ | 3116 | 2271 | 3091 | 2251 | 3030 | 2197 |
| QZ/DZ | $4 \times 10^{-5}$ | $5 \times 10^{-7}$ | 2700 | 1836 | 2636 | 1767 | 2458 | 1587 |

**Table 2:** Multicomponent HCI and "fixed" FGH protonic excited-state energies for the HCN and FHF- systems. All energies are in $cm^{-1}$ and are relative to the ground-state energy (n_state=1). An n_state of 2 or 3 corresponds to a degenerate bending mode and an n_state of 4 corresponds to the stretching mode.

From Table 2, it is seen the inclusion of the PT2 correction increases the accuracy of the calculated excited states significantly. The change in the calculated excited state values is greater for electronic basis set combinations with more electronic basis functions. This is mainly a consequence of the difference between the exact CI and variational energies being larger for basis set combination with more electronic basis functions. For the DZ/DZ electronic basis set combination, the variational calculations are already able to approach the exact CI limit, so the perturbative correction has less of an effect.

Examining the QZ/DZ electronic basis set combination, which is the largest basis set combination that protonic excitation calculations were performed with all the multicomponent HCI, EOM-CCSD, and TDDFT methods, we see that multicomponent HCI has a smaller absolute error for all protonic excitation values compared to the multicomponent EOM-CCSD method. Multicomponent HCI has a larger absolute error than multicomponent TDDFT for both HCN protonic excitation energies and the FHF$^-$ bending excitation energy. Multicomponent HCI has a smaller absolute error than multicomponent TDDFT for the FHF$^-$ stretching excitation energy. We note that for the multicomponent TDDFT method we compare to DFT FGH protonic excitation values, while for multicomponent HCI, we compare to the CCSD FGH protonic excitation values. Overall, given the results in this study, the multicomponent HCI method appears to be competitive with other state-of-the-art methods for calculating protonic excitation energies.

## C. Extrapolation of Energies

In previous single-component HCI studies, the perturbative correction enabled the implementation of an extrapolation procedure to obtain estimates of the total energy with $\varepsilon_{VAR} = 0$.[39, 41] This is done by performing a series of HCI calculations with different values of $\varepsilon_{VAR}$ with $\varepsilon_{PT2}$ sufficiently small that the energy is converged with respect to $\varepsilon_{PT2}$ for each value of $\varepsilon_{VAR}$. A graph of the total energy, $E_{tot,}$ versus the difference in the variational energy and total energy, $E_{var}$-$E_{tot}$, gives a linear relationship. Using this line, extrapolation can be performed to the $E_{var}$-$E_{tot}$ = 0 limit for the total energy to obtain an estimate of $E_{TOT}$ at the $\varepsilon_{VAR} = 0$ limit. We have performed such an extrapolation for the HCN and FHF$^-$ molecules with the TZ/DZ and QZ/DZ electronic basis set combinations, which are shown in Figures 4 and 5, respectively.

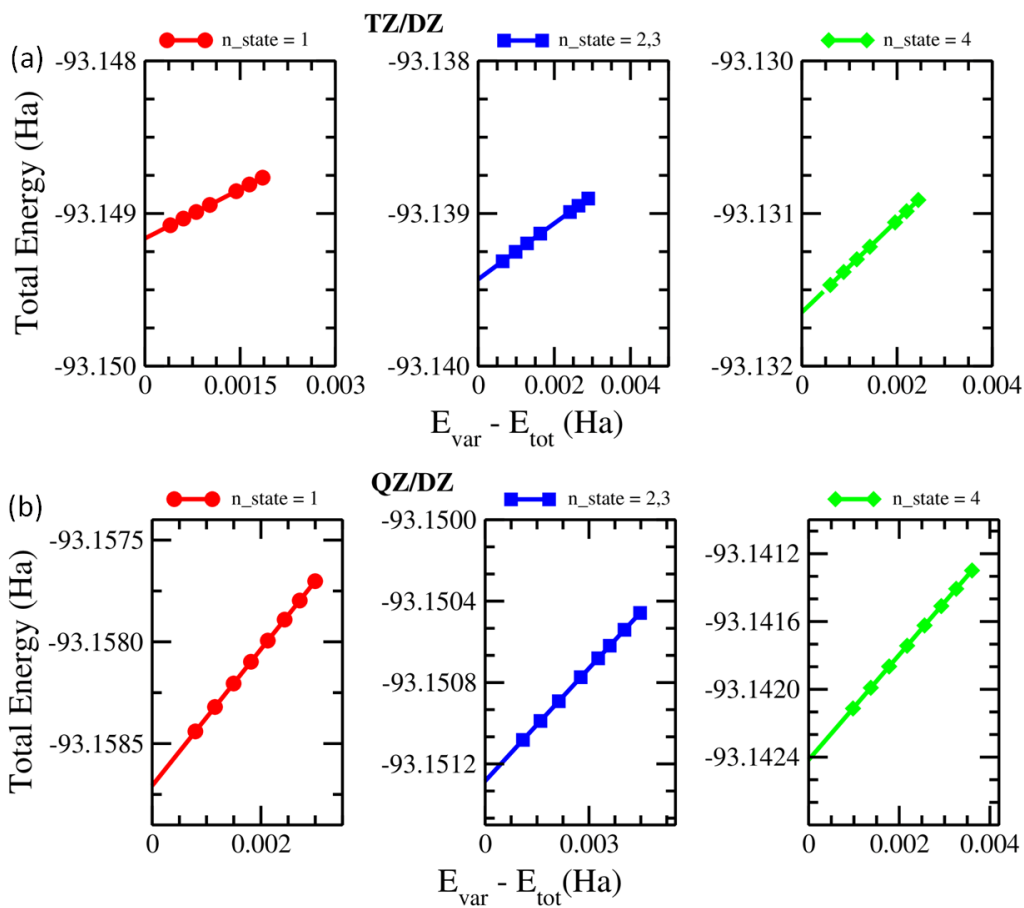

**Figure 4:** Extrapolated energy curves for the ground state and excited states of HCN. The lowest $\varepsilon_{VAR}$ values considered for TZ/DZ and QZ/DZ basis sets are $1.5 \times 10^{-5}$ and $3.0 \times 10^{-5}$, respectively. An n_state of 2 or 3 corresponds to a degenerate bending mode and an n_state of 4 corresponds to the stretching mode.

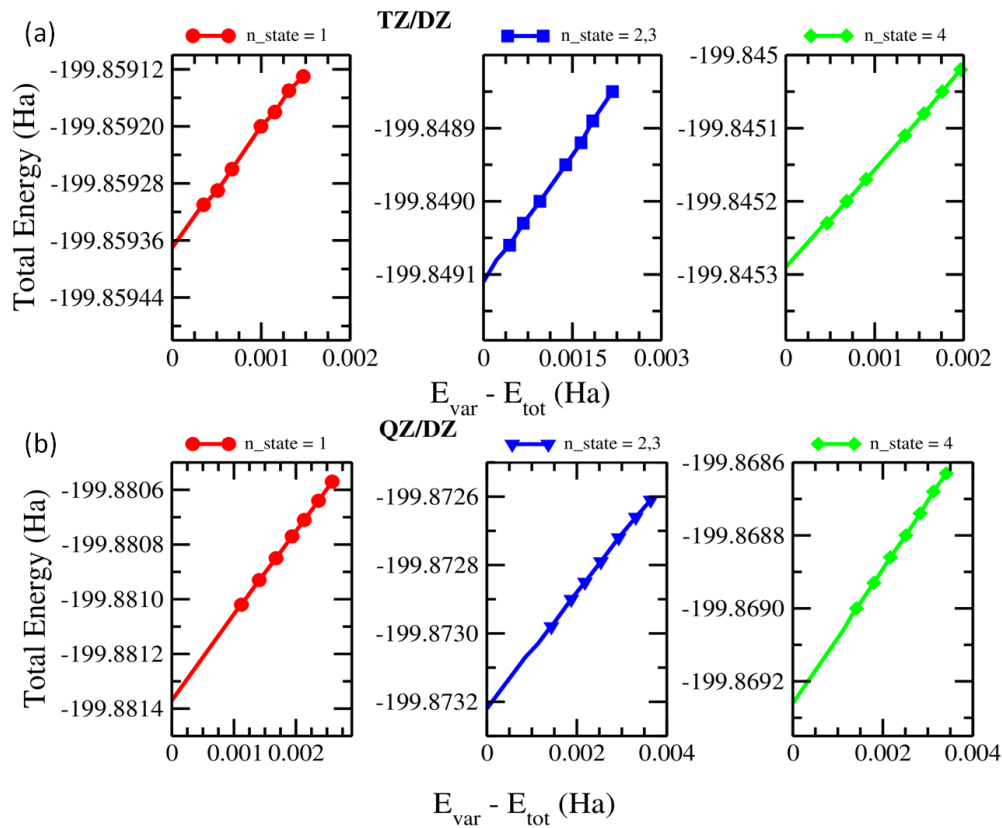

Figure 5: Extrapolated energy curves for the ground state and excited states of FHF⁻. The lowest $\varepsilon_{VAR}$ values considered for TZ and QZ basis sets are $3.0 \times 10^{-5}$ and $4.0 \times 10^{-5}$, respectively. An n_state of 2 or 3 corresponds to a degenerate bending mode and an n_state of 4 corresponds to the stretching mode.

From these Figures, there is a clear linear relationship between $E_{TOT}$ and $E_{var}$-$E_{tot}$ for all plots. We have fit a line to each of these graphs to obtain an extrapolated estimation of the energy for each of the states. Using these values, we calculated the excitation energies again, which are presented in Table 3. We see that the extrapolation has only a very small effect on the calculated protonic excitation energies, with the largest change being 22 cm$^{-1}$. This is largely due to the slopes of all the lines in the extrapolations being similar. I.e., the extrapolation procedure changes the energies of all states by similar amounts. More testing is needed to see if this trend is true in general for multicomponent protonic excitation energies.

We note that the extrapolation procedure itself results in a non-negligible change for the ground-state energy of the FHF$^-$ molecule of approximately 0.4 mHa such that when calculating absolute energies or when the energy of a multicomponent system that has undergone a reaction or rearrangement of the atoms, the change from the extrapolated energy is likely to be more significant.

| | HCN | | | | FHF⁻ | | | |
|---|---|---|---|---|---|---|---|---|
| | HCl | | FGH | | HCl | | FGH | |
| n_state | 4 | 2,3 | 4 | 2,3 | 4 | 2,3 | 4 | 2,3 |
| TZ/DZ | 3844 | 2135 | 3440 | 1928 | 3090 | 2252 | 3030 | 2197 |
| QZ/DZ | 3575 | 1629 | 3202 | 1076 | 2658 | 1788 | 2458 | 1587 |

**Table 3:** Extrapolated multicomponent HCl and FGH protonic excited-state energies for the HCN and FHF⁻ molecules. All energies are in $cm^{-1}$ and are relative to the ground-state energy (n_state=1). An n_state of 2 or 3 corresponds to a degenerate bending mode and an n_state of 4 corresponds to the stretching mode.

## V. Conclusions

A new procedure for generating reference FGH protonic excitation energies has been introduced where the electronic basis functions of the quantized particle are fixed at the position from a Born-Oppenheimer geometry optimization. For a given one-particle electronic basis set, the FGH results agree much better with previous multicomponent protonic excitation energy calculations and the results in the study should be useful in future studies that benchmark new multicomponent methods as the results should allow the new methods to be tested without the need for large electronic basis sets centered on the quantized particle.

The multicomponent HCI method has been extended to the calculation of protonic excited states and the PT2 correction for multicomponent HCI has been derived and implemented. The multicomponent HCI method with the PT2 correction is shown to perform similarly compared to other leading multicomponent methods for calculating protonic excitations and the PT2 correction is shown to lower the errors in the excitation energies relative to reference FGH calculations. Given the CI-based formalism of multicomponent HCI, it is likely to be conceptually simple to extend it to the calculation of one-electron one-proton excitations. The accurate calculation of such excitations has been a long-standing goal for multicomponent methods.[55]

Finally, the PT2 correction could in principle be included directly in multicomponent complete active space self-consistent field calculations[74] in a manner identical to previous single-component HCI studies.[31] For single-component orbital-optimized selected CI methods, there has been debate[34] about whether the PT2 correction being included in the orbital-optimization procedure offers any benefit, but given the success of orbital-optimized multicomponent methods relative to their standard multicomponent counterparts, such inclusion is likely to work better in the multicomponent rather than single-component case.

## Acknowledgements

KRB thanks the University of Missouri-Columbia for start-up funding.

## Author Declarations

The authors have no conflicts to disclose.

## AIP Publishing Data Sharing Policy

The data that support the findings of this study are available within the article.

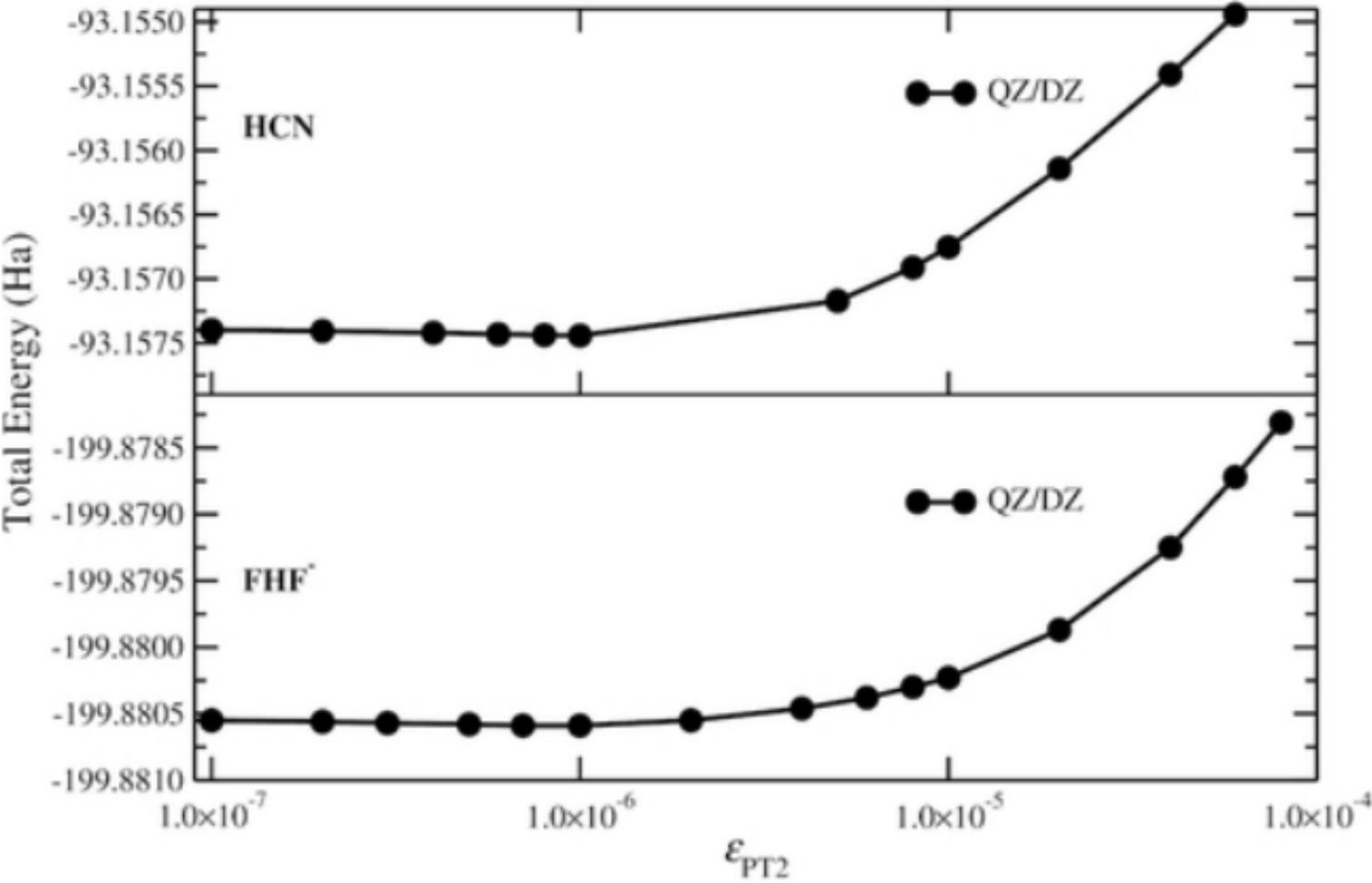

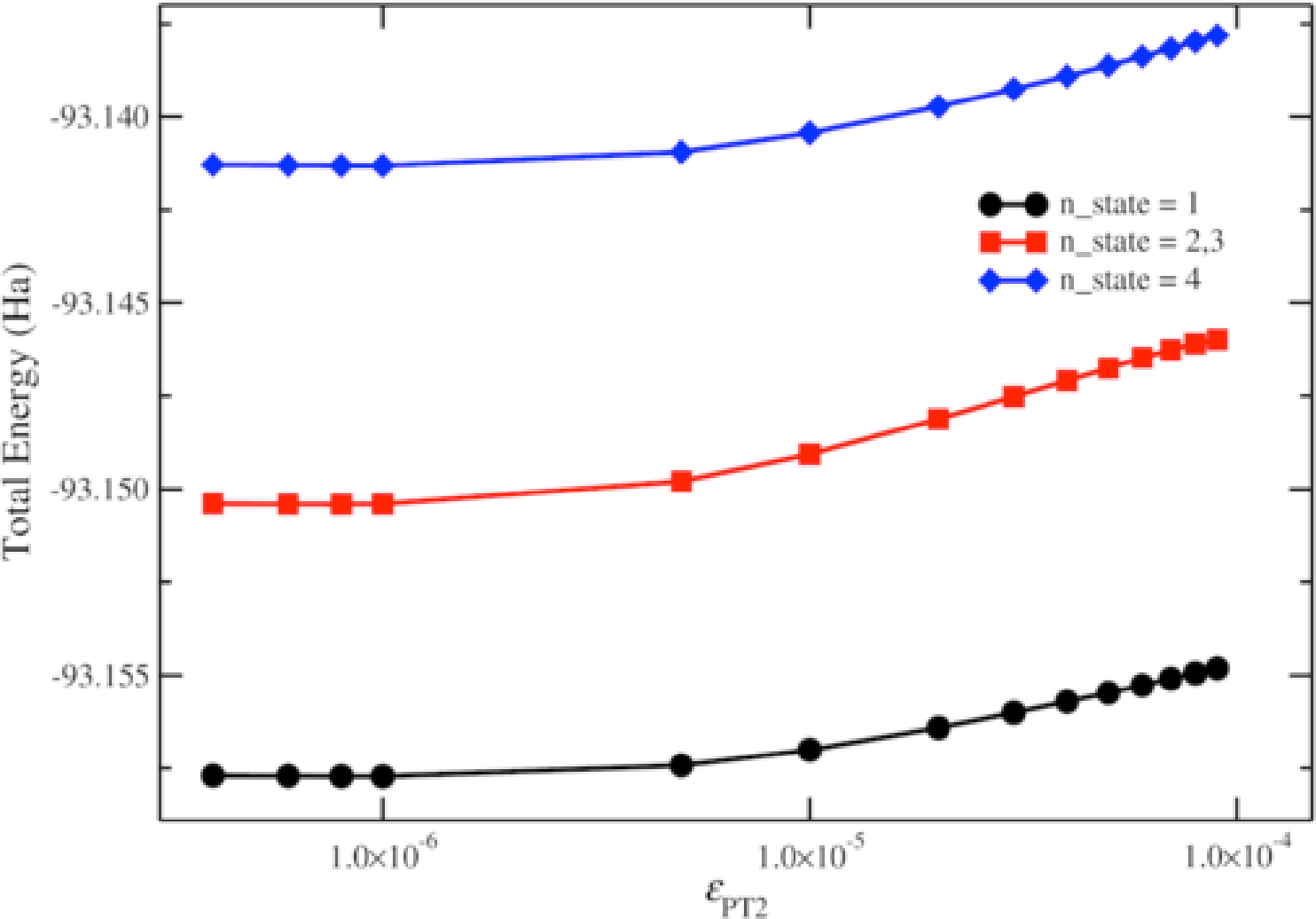

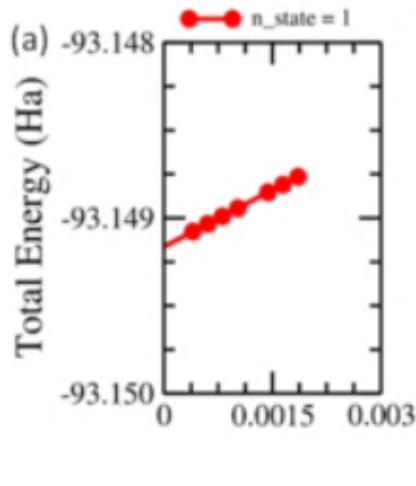
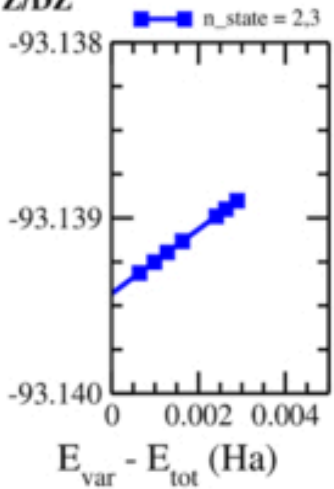
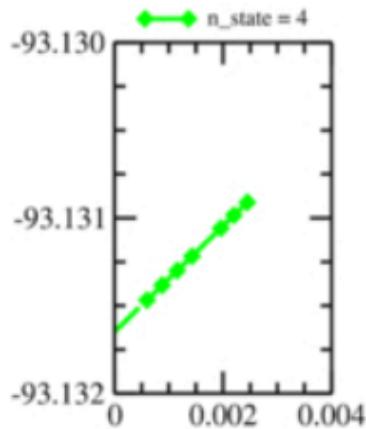
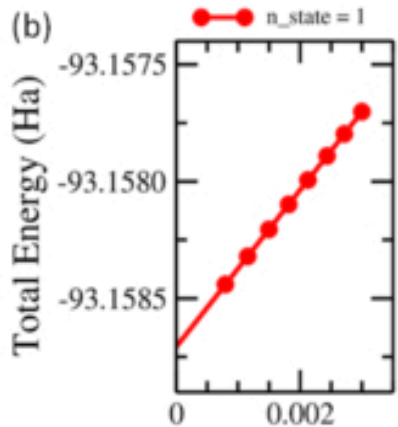
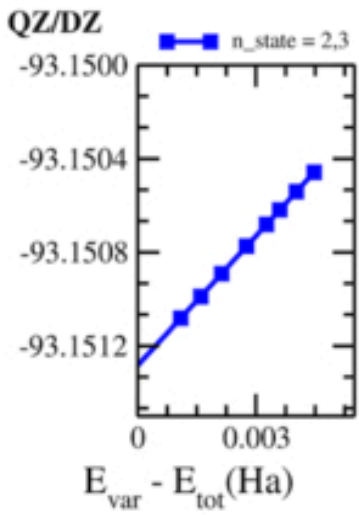
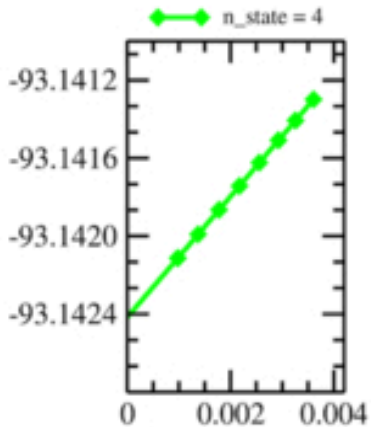

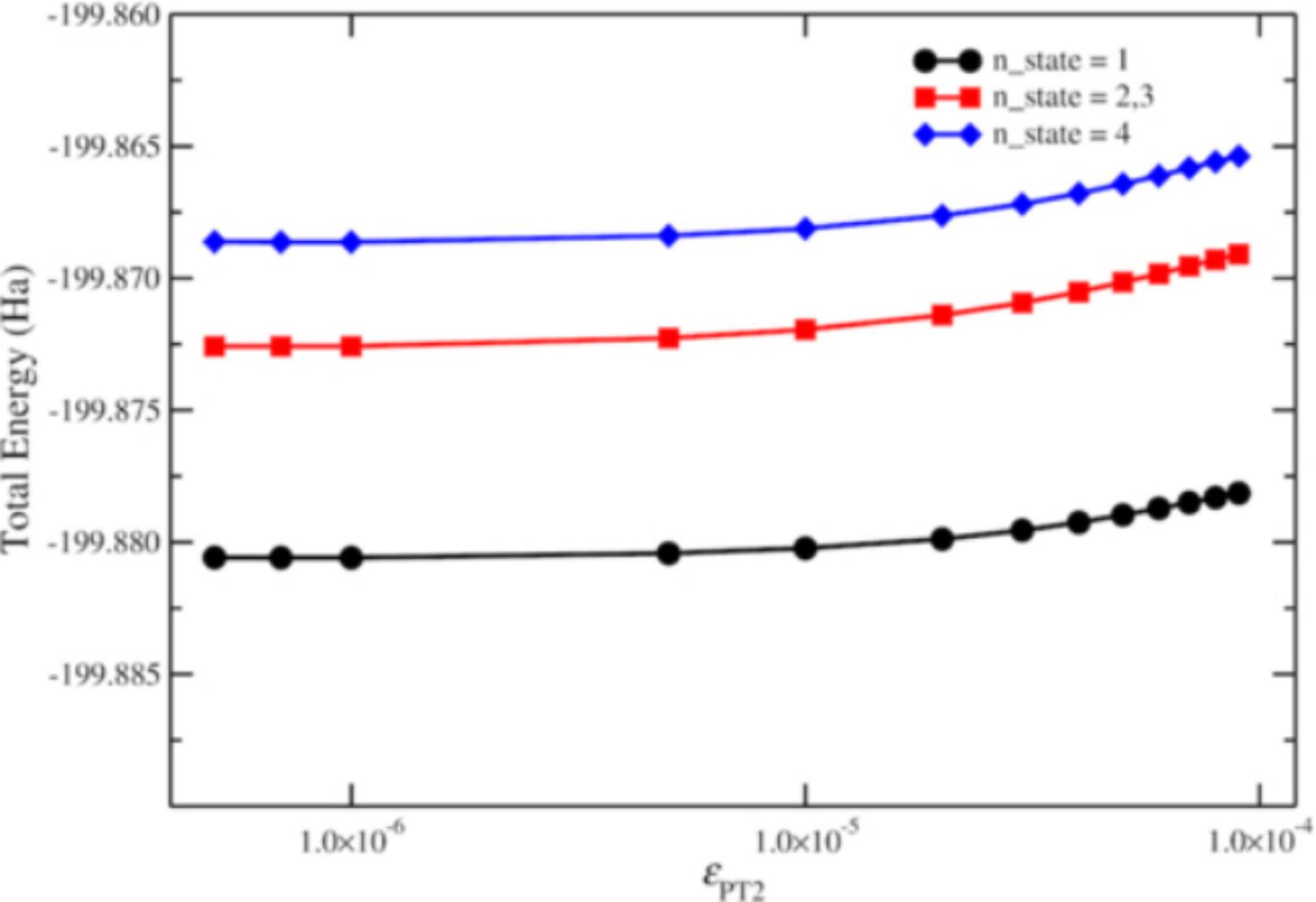

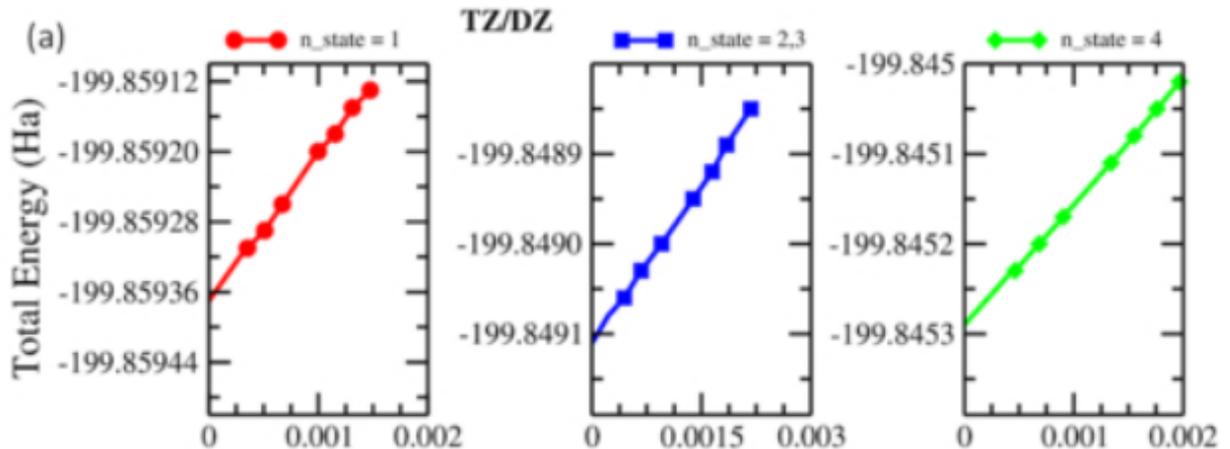

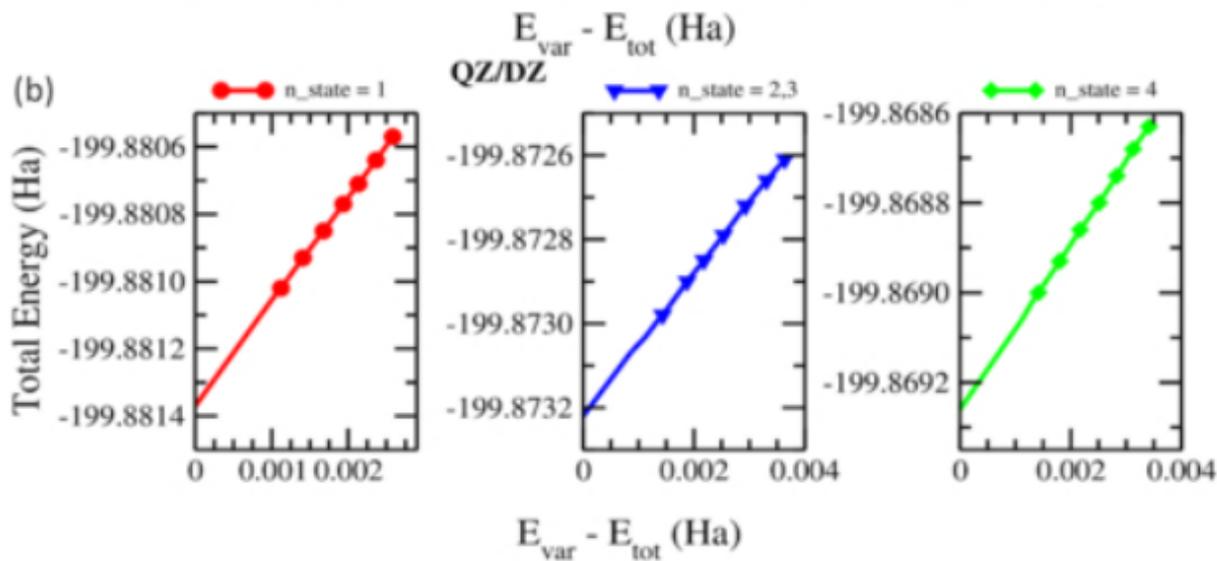